\documentclass[prl,reprint]{revtex4-2}

\usepackage[english]{babel}
\usepackage{graphicx}
\usepackage{amsmath}
\usepackage{amsfonts}
\usepackage[utf8]{inputenc}
\usepackage{xcolor}
\usepackage{physics}
\usepackage{overpic,color}
\usepackage{hyperref}
\usepackage{dsfont}
\usepackage{mathtools}
\usepackage{bm}
\usepackage[normalem]{ulem}
\usepackage{mathtools}

\newcommand{\hamiltonianglobal}{\hat{H}}
\newcommand{\hamiltoniansystem}{\hat{H}_{\text{S}}}
\newcommand{\hamiltonianclock}{\hat{H}_{\text{C}}}
\newcommand{\energyglobal}{E}
\newcommand{\energyclock}{E_{\text{C}}}
\newcommand{\interactionfunction}{V}
\newcommand{\interaction}{\hat{\interactionfunction}}
\newcommand{\potentialsystem}{\interaction_{\text{S}}}
\newcommand{\stateglobal}{\Psi}
\newcommand{\statesystem}{\varphi}
\newcommand{\stateclock}{\chi}
\newcommand{\ketstateglobal}{\ketglobal{\stateglobal}}
\newcommand{\prm}{\lambda}
\newcommand{\hilbert}{\mathcal{H}}
\newcommand{\hilbertsystem}{\hilbert_{\text{S}}}
\newcommand{\hilbertclock}{\hilbert_{\text{C}}}
\newcommand{\ketsys}[1]{\ket{#1}_{\text{S}}}
\newcommand{\ketclock}[1]{\ket{#1}}
\newcommand{\ketglobal}[1]{\ket*{#1}\!\rangle}
\newcommand{\braclock}[1]{\bra{#1}}
\newcommand{\braketclock}[2]{\braket{#1}{#2}}
\newcommand{\braketclockglobal}[2]{\braket*{#1}{#2}\!\rangle}
\newcommand{\melclock}[3]{\mel{#1}{#2}{#3}}
\newcommand{\melclockglobal}[3]{\mel*{#1}{#2}{#3}\!\rangle}
\newcommand{\melglobal}[3]{\mel*{\!\langle #1}{#2}{#3 \rangle\!}}
\newcommand{\braketglobal}[2]{\braket{\!\langle #1}{#2 \rangle\!}}
\newcommand{\ident}{\hat{\mathds{1}}}
\newcommand{\identsystem}{\ident_{\text{S}}}
\newcommand{\identclock}{\ident_{\text{C}}}
\newcommand{\dyadclock}[1]{\dyad{#1}}
\newcommand{\dyadglobal}[1]{|{#1}\rangle\!\rangle\hspace{-0.5mm}\langle\!\langle{#1}|}
\newcommand{\clockoperator}[1]{\hat{#1}_{\text{C}}}
\newcommand{\projectorglobal}{\hat{P}_{\stateglobal}}
\newcommand{\projectorclock}{\hat{P}_{\stateclock}}
\newcommand{\projectorconstrained}{\hat{P}_{\stateglobal\stateclock}}
\newcommand{\scalarpotential}{\mathcal{E}}
\newcommand{\meanval}[1]{\mathsf{#1}}
\newcommand{\meanvalclock}[1]{\mathsf{#1}_{\text{C}}}
\renewcommand{\vec}[1]{\mathbf{#1}}
\newcommand{\unitary}{\hat{U}}
\newcommand{\unitarysystem}{\unitary_{\text{S}}}
\newcommand{\unitaryclock}{\unitary_{\text{C}}}
\newcommand{\sigmax}{\hat{\sigma}_x}
\newcommand{\sigmay}{\hat{\sigma}_y}
\newcommand{\sigmaz}{\hat{\sigma}_z}
\newcommand{\sigmaxclock}{\hat{\sigma}_{\text{C},x}}

\newcommand{\sigmazclock}{\hat{\sigma}_{\text{C},z}}
\newcommand{\sigmaxsystem}{\hat{\sigma}_{\text{S},x}}
\newcommand{\sigmaysystem}{\hat{\sigma}_{\text{S},y}}
\newcommand{\sigmazsystem}{\hat{\sigma}_{\text{S},z}}
\newcommand{\sigmavecsystem}{\hat{\bm{\sigma}}_{\text{S}}}
\newcommand{\up}{\uparrow}
\newcommand{\down}{\downarrow}
\newcommand{\upclock}{\up_{\text{C}}}
\newcommand{\downclock}{\down_{\text{C}}}
\newcommand{\upsystem}{\up_{\text{S}}}
\newcommand{\downsystem}{\down_{\text{S}}}

\newcommand{\changes}[1]{\textcolor{black}{#1}}

\begin{document}

\title{The emergence of time from quantum interaction with the environment}

\author{Sebastian Gemsheim}
\email{sebgem@protonmail.com}
\author{Jan M. Rost}
\affiliation{Max Planck Institute for the Physics of Complex Systems, N\"othnitzer Stra\ss e 38, D-01187 Dresden, Germany}

\begin{abstract}
The nature of time as emergent for a system by separating it from its environment has been put forward by Page and Wootters [D. N. Page and W. K. Wootters, Phys. Rev. D 27, 2885 (1983)] in a quantum mechanical setting neglecting interaction between system and environment. Here, we add strong support to the relational concept of time by deriving the time-dependent Schr\"odinger equation for a system  from an energy eigenstate of the global Hamiltonian consisting of system, environment \emph{and} their interaction. Our results are consistent with concepts for the emergence of time where interaction has been taken into account at the expense of a semiclassical treatment of the environment. Including the coupling between system and environment without approximation adds a missing link to the relational time approach opening it to dynamical  phenomena of interacting systems and entangled quantum states.
\end{abstract}

\maketitle


The nature and role of time to decipher the physical world is a basic and persisting research topic, in particular the question, if time is fundamental or emergent. 
For the latter, the starting point is a static  description of the world. Time emerges from  singling out a system from the rest of the world, its environment. As such, time is a meaningful tool to describe the relation of system and environment, both governed by Hamiltonians  distinguished in physical or abstract (Hilbert) space. This has lead to two strands of research for the relational approach to time. One strand, initiated by Page and Wootters~\cite{Page1983, Wootters1984, Giovannetti2015, Marletto2017, Hoehn2021} deals with abstract state vectors in Hilbert space and is analytically exact, but remains to date  unable to  deal with general couplings of system and environment. The second strand uses a semiclassical approach typically in position space, arguing that the environment is ``large enough" to allow for semiclassical approximations~\cite{Halliwell1985, Kuchar1992, Kiefer1994, Anderson2012, Kiefer2012, Briggs2000, Briggs2001, Schild2018, Chataignier2022}. By these means, time also emerges as relation between system and environment which may be arbitrarily coupled.

Here, we will show how time emerges  quantum mechanically in the relation between system and environment without approximations, more specifically, by retaining arbitrary couplings between them and without the need to resort to semiclassical approximations.
That is, starting from a static  global state encompassing system and environment we derive the time-dependent Schr\"odinger equation including an arbitrary, time-dependent potential for the system in a few transparent steps. To this end, we will  re-formulate the stationary (timeless) Schr\"odinger equation for the global state as an \textit{invariance principle} and single out a pure state of the system from its inevitable embedding in the environment by projecting a specific state of the environment onto the global state. 
As a by-product our approach constitutes a concept for analytical solutions of complicated time-dependent interaction potentials~\cite{Gemsheim2023}.

\begin{figure}
    \centering
    \includegraphics[width=0.9\columnwidth]{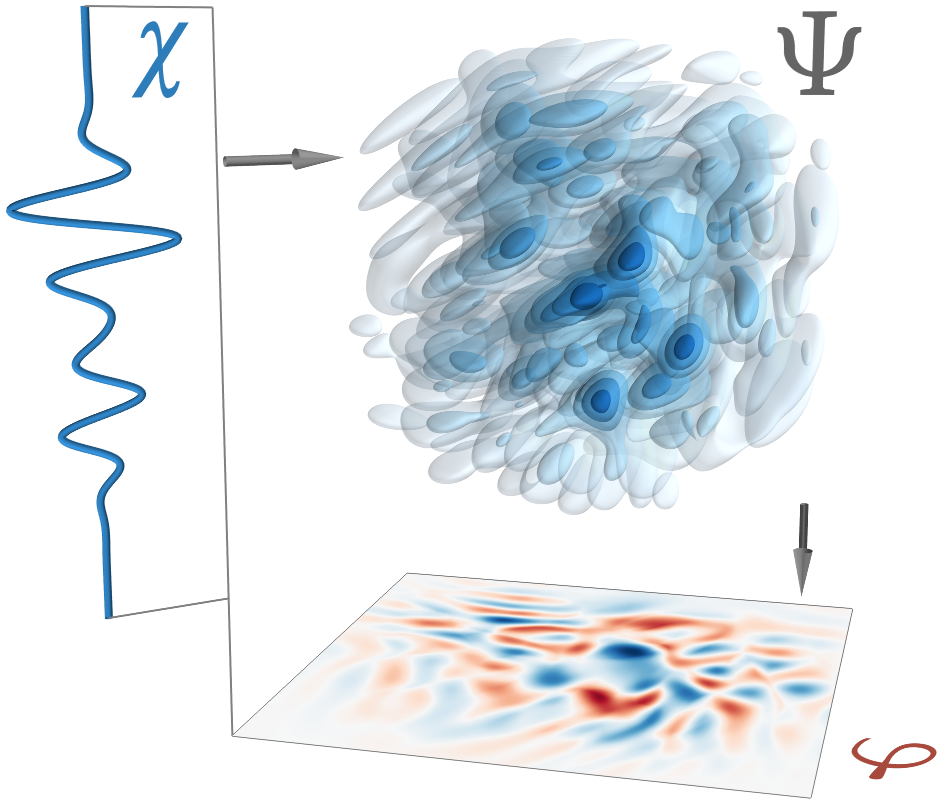}
    \caption{Sketch of the relational state formalism. A one-dimensional environment state $\stateclock(x)$ projects out a two-dimensional system state $\statesystem(y,z) \propto \int \dd{x} \stateclock^*(x) \stateglobal(x,y,z)$ from the three-dimensional global state $\stateglobal(x,y,z)$. Schematically, the clock wavefunction is multiplied to each vertical column of $\stateglobal$ and subsequently integrated along this direction to yield each value of $\statesystem$. With such an inherent clock dependence, the system state generally differs for different clock states. 
    }
    \label{fig:conditional_state}
\end{figure}

The invariance principle for the global state $\ketstateglobal$ as an eigenstate of the Hamiltonian $\hamiltonianglobal$ with global eigenenergy $\energyglobal$ reads
    \begin{equation}
        \exp \left[ i \prm ( \hamiltonianglobal - \energyglobal ) \right] \ketstateglobal 
        = \ketstateglobal
        \label{eq:global_invariance}
    \end{equation}
for all complex $\prm$  with dimension of inverse energy, where $\braketglobal{.}{.}$ stands for the scalar product in the global Hilbert space.
Differentiating \eqref{eq:global_invariance} w.r.t.~$\lambda$ gives the (timeless) Schr\"odinger equation $( \hamiltonianglobal - \energyglobal ) \ketstateglobal = 0 $, often referred to as TISE.
In the following, we will only consider real-valued $\lambda$ in \eqref{eq:global_invariance} which is sufficient to demonstrate  the emergence of time. Purely imaginary $\lambda$ finds its natural application in the emergence of temperature~\cite{Gemsheim2023b}.
In order to single out a system state from the global state, we first partition the global Hamiltonian $\hamiltonianglobal$ into that of the system $\hamiltoniansystem$, its environment $\hamiltonianclock$  and their possible interaction $\interaction$,
    \begin{equation}
        \hamiltonianglobal 
        = \hamiltoniansystem \otimes \identclock + \identsystem \otimes \hamiltonianclock + \interaction\,.
      \label{eq:hamiltonian_sum}
    \end{equation}
We will use environment  and clock as synonyms to relate to the aforementioned two strands of research on the emergence of time.
While the partition \eqref{eq:hamiltonian_sum} of the global Hamiltonian  is natural to define a system in the first place, it is not  obvious how to single out a system state from the global, \emph{entangled} state $\ketglobal{\stateglobal}$. From a quantum mechanical point of view, the system is inevitably embedded in its environment on which it is therefore conditioned. Hence, a system state $\ketsys{\statesystem}$ is created by projecting the global state onto a state of the environment, $\ketsys\statesystem =  \braketclockglobal{\stateclock}{\stateglobal}$~\footnote{Since the clock projection includes an identity operation on the system part, it is fully described by $(\identsystem \otimes \braclock{\stateclock}) \ketstateglobal    = \sum_{kl} c_{kl} (\identsystem \ketsys{\statesystem_k} ) \otimes \braketclock{\stateclock}{\stateclock_l} = \sum_k \ketsys{\statesystem_k} (\sum_l c_{kl} \braketclock{\stateclock}{\stateclock_l}) \in \hilbertsystem $ for a representation of $\ketstateglobal$ in the arbitrary global basis $\{ \ketglobal{\statesystem_k \otimes \stateclock_l} \}_{kl}$. We will omit those  identity operators where obvious in favor of a better readability.}.
Here and in the following we use the convention that $\braket{.}{.}$ and $\braketglobal{.}{.}$ denote scalar products in environment and full Hilbert space, respectively, while $\ketsys\statesystem$ and $\ket\stateclock$ stand for states of system and environment, respectively and $\ketstateglobal$ is reserved for the global state. A sketch of this relational approach is shown in Fig.~\ref{fig:conditional_state}.

Singling out the system by projection reduces the correlations and in particular breaks the global symmetry such that  the system state does not obey the global invariance principle. Rather, the state becomes dependent on the symmetry parameter $\prm$.
This can be seen by projecting the invariance equation \eqref{eq:global_invariance} onto $\bra{\stateclock_0}$, which gives for the interaction free case, $V=0$,
\begin{equation}
    \melclockglobal{\stateclock_0}{ e^{i\prm(\hamiltonianclock - \energyglobal)} }{\stateglobal} 
    = e^{-i\prm \hamiltoniansystem} \braketclockglobal{\stateclock_0}{\stateglobal} \,,
    \label{eq:projected-invariance-V0}
\end{equation}
where we may write
\begin{equation}
    \ketclock{\stateclock_{\lambda}} = e^{-i\prm(\hamiltonianclock - \energyglobal)}  \ketclock{\stateclock_0} \equiv
    \unitaryclock(\prm) \ketclock{\stateclock_0}\,.
    \label{eq:clock_state_evolved}
\end{equation}
The  states $\ketclock{\stateclock_\prm}$ from the environment serve as markers to tag the system state with $\lambda$,
\begin{equation}
    \ketsys{\statesystem(\prm)}\equiv  
     \braketclockglobal{\stateclock_\lambda}{\stateglobal} \,.
    \label{eq:system_state_evolved_definition}
\end{equation}
Consistent with $ \ketsys{\statesystem(0)}=\braketclock{\stateclock_0}{\stateglobal}$, we arrive at
\begin{equation}
    \ketsys{\statesystem(\prm)} 
    =  e^{-i\prm \hamiltoniansystem} \ketsys{\statesystem(0)} \equiv \unitarysystem(\prm)\ketsys{\statesystem(0)}
    \label{eq:system_state_evolved}
\end{equation}
for all symmetry parameters $\prm$. Hence, we have \textit{derived} from the global invariance  \eqref{eq:global_invariance} without reference to any differential equations how states of the system \eqref{eq:system_state_evolved} and the environment \eqref{eq:clock_state_evolved} evolve. This implies a peculiar consequence  on the fundamental level: states with different $\prm$ do not have to be related, admitting also discrete symmetries  with $\prm$ replaced by a set of parameters $\{\prm_n\}$. 

Using the property $\unitary^\dagger(\prm)=\unitary(-\prm)$ of the unitary transformations in \eqref{eq:clock_state_evolved} and \eqref{eq:system_state_evolved} we can rewrite the projected invariance equation \eqref{eq:projected-invariance-V0} as
\begin{align}
    \braketclockglobal{\stateclock_0}{\stateglobal} &=  \unitarysystem(-\prm)\melclockglobal{\stateclock_0}{\unitaryclock(-\prm)}{\stateglobal}\nonumber\\
     &= \unitarysystem(-\prm)\braketclockglobal{\unitaryclock(\prm)\stateclock_0}{\stateglobal} \,,
     \label{eq:projected2TISE_no_interaction}
\end{align}
which has the same form as the invariance for more familiar symmetry transformations, e.g., the invariance of  a state $\ket\psi$ in coordinate space  $\braket{\vec r}{\psi}$ if it  is rotated by an angle $\theta$ about a vector $\vec u$ with the unitary operator $\hat{D}(\theta)=e^{-i\theta \vec u\cdot\hat{\vec J}/\hbar}$ while the coordinate system is rotated backwards with the rotation matrix $R(\theta)$: $\hat{D}(\theta)\braket{R(-\theta)\vec{r}}{\psi} = \braket{\vec{r}}{\psi}$. This opposite behavior of states of the system and environment as a consequence of the global invariance was dubbed by Zurek ``envariance'' and used to motivate, why probabilities correspond to measurements, colloquially known as the Born Rule \cite{Zurek03}.

In our context of letting time emerge by  projection of a globally static state, we may conclude that for the projected global invariance \eqref{eq:projected-invariance-V0} the state $\ketclock{\stateclock}$ from the environment plays the role of a coordinate which is  transformed with $\unitaryclock(\prm)$ to compensate the transformation of the system state $\ketsys{\statesystem}$  with $\unitarysystem(-\prm)$. 

Since $\prm$   in \eqref{eq:global_invariance} is a continuous symmetry, \eqref{eq:system_state_evolved} can be interpreted as the solution of the differential equation
\begin{equation}
    i \dv{\prm} \ketsys{\statesystem(\prm)} 
    = \hamiltoniansystem \ketsys{\statesystem(\prm)}\,
     \label{eq:TDSE}
\end{equation}
with initial condition $ \ketsys{\statesystem(0)}= \braketclockglobal{\stateclock_0}{\stateglobal}$.
Obviously, \eqref{eq:TDSE} is equivalent to the TDSE if time $t$ is introduced  through  $\lambda =  t/\hbar$. 
What we have described so far is a short cut derivation of the Page-Wootters relational time approach \cite{Page1983} made possible by recognizing the crucial role of the invariance principle \eqref{eq:global_invariance}.

Strictly speaking, $\prm$ is  only a label without physical meaning: Any re-parametrization  ${\prm} = f(\tilde\prm)$ leaves the relations between environment  and system invariant. 
However, one can tag the system's evolution with a reparametrization invariant  observable of the environment, $\meanvalclock{A}(\prm) \equiv \melclock{\stateclock_\prm}{\clockoperator{A}}{\stateclock_\prm} : \hilbertclock \mapsto \mathds{R}$.  Although $\clockoperator{A}$ operating on the environment is   arbitrary  apart from being Hermitian, a good choice is one for which the relation between $\prm$ and $\meanvalclock{A}$ is simple, for example linear, if the environment is used as a clock. This idea goes back to Poincar{\'e}~\cite{Callender2017}. For instance, the mean position $\meanval{R}(\prm) = \prm\, \meanval{P}(0)/M + \meanval{R}(0)$ of a free particle of mass $M$ with $\hamiltonianclock = \hat{P}^2 / 2M$ can reliably track dynamics for non-vanishing  mean momentum $\meanval{P}(0) \neq 0$ since we can  replace $\prm = M[ \meanval{R}(\prm)-\meanval{R}(0)]/\meanval{P}(0)$ which represents a physical property of the environment, respectively clock.
For a  state $\ketclock{\stateclock_{\prm}}$ to clock the system, it must first of all have overlap with the global state  (see Fig. 1). To provide a high resolution in $\prm$, the clock state $\ketclock{\stateclock_{\prm}}\propto  \sum_k a_k e^{-i\prm E_{C,k}} \ketclock{E_{C,k}}$ must be distributed over many eigenstates $\ketclock{E_{C,k}}$  of $\hamiltonianclock$, with ideally $|a_k| \approx \text{const}$  \cite{Giovannetti2015, Marletto2017, Smith2019}. This is easy to realize, if the (physical) dimension of the clock is much larger than that of the system, which  also has the effect that the global state can accommodate more complex system dynamics.

We also re-emphasize that the  entanglement in $\ketstateglobal$ with respect to the states of system and environment is crucial for non-trivial system dynamics and requires without interaction $\interaction$ the existence of degenerate eigenspaces of the global Hamiltonian. 
Otherwise, system and environment fulfill separately a ``global" invariance principle with  $\prm_{\text{S}}$ and $\prm_{\text{C}}$, respectively, which leaves the relation $\prm_{\text{S}}(\prm_{\text{C}})$ undetermined.

Finally, it is remarkable that despite the global invariance having been broken by an arbitrary but specific choice of $\ket{\stateclock_0}$, the properties of the latter do not influence the evolution of the system state other than specifying its initial condition.  Hence, the standard procedure of getting rid of properties of the environment  to achieve a universal system evolution, namely tracing over the environment, is not necessary.
While it is contained in the present description (we could use any kind of mixed state for  $\ket{\stateclock_0}$), choosing a rather structureless $\ket{\stateclock_0}$ is not suitable for serving the purpose of a clock  as just discussed.

So far we have provided a clarification and short-cut to the TDSE for a system not interacting  with its environment, enabled by recognizing the power of the invariance principle \eqref{eq:global_invariance} which was not invoked in \cite{Page1983}. We have detailed our approach since we need it in the following to derive  the TDSE for a system interacting with the environment.
    
In reality, the environment, will inevitably interact with the system. This automatically ensures that the global state $\ketstateglobal$ is generically entangled. 
Hence, we should derive the TDSE for the system with interaction $\interaction\ne0$. 
To this end, we use  $\ketclock{\stateclock(\prm)} = e^{-i S(\prm)}\ketclock{\stateclock_\prm}$ with $\ketclock{\stateclock_\prm}$ from \eqref{eq:clock_state_evolved} and the complex scalar $S(\prm)=\int^\prm d\prm' \scalarpotential(\prm')$, which can be viewed as a $\prm-$dependent phase and normalization. Projected onto this state,
the global TISE can be written as 
\begin{equation}
    \left(-\hamiltoniansystem + \scalarpotential(\prm)+ i\dv{\prm}\right )\braketclockglobal{\stateclock(\prm)}{\stateglobal}=\melclockglobal{\stateclock(\prm)}{\interaction}{\stateglobal}\,.
    \label{eq:projected2TISE}
\end{equation}
As a next step we decompose $\melclockglobal{\stateclock(\prm)}{\interaction}{\stateglobal}$ into a Hermitian potential $\potentialsystem(\prm)$ for the system and a c-number which is an expectation value over the global state. The decomposition is facilitated with the operators $\projectorglobal \equiv \dyadglobal{\stateglobal}$, $\projectorclock \equiv \identsystem \otimes \dyadclock{\stateclock(\prm)}$ and $\projectorconstrained = \projectorglobal\projectorclock/N_\prm$, where $\projectorconstrained\ketstateglobal=\ketstateglobal$ since $N_\prm = \melglobal{\stateglobal}{\projectorclock}{\stateglobal}$.
We obtain 
 \begin{subequations}
 \label{eq:decomposition}
\begin{align}
    \melclockglobal{\stateclock}{\interaction}{\stateglobal} 
    &= \melclockglobal{\stateclock}{\interaction\projectorconstrained}{\stateglobal}\nonumber \\
    &= \left[ \potentialsystem(\prm) - \melglobal{\stateglobal}{ \interaction \projectorclock }{\stateglobal}/N_{\prm} \right] \braketclockglobal{\stateclock(\prm)}{\stateglobal}
\end{align}
where
\begin{equation}
 \label{eq:effective_potential}
      \potentialsystem(\prm)  =\frac{
    \melclock{\stateclock}{ \left(\interaction\projectorglobal + \projectorglobal\interaction \right) }{\stateclock}}{\melglobal{\stateglobal}{\projectorclock}{\stateglobal} } \,.
\end{equation}
\end{subequations}
Inserting \eqref{eq:decomposition} into \eqref{eq:projected2TISE}, setting $\scalarpotential(\prm) {\equiv} \melglobal{\stateglobal}{\interaction\projectorclock}{\stateglobal}/N_\prm$ and rearranging terms gives the TDSE for the system with interaction,
 \begin{equation}
 \left[ \hamiltoniansystem + \potentialsystem(\prm) \right] \ketsys{\statesystem(\prm)} =
    i \dv{\prm} \ketsys{\statesystem(\prm)} \,.
    \label{eq:tdse_with_effective_potential}
\end{equation}
The effective system potential $\potentialsystem$ from \eqref{eq:effective_potential} depends explicitly on $\prm$  and implicitly on the state  of the environment, $\ketclock{\stateclock(\prm)}=e^{-i\prm ( \hamiltonianclock - \energyglobal )-i S(\prm)}\ketclock{\stateclock_0} $.
One can easily retrieve the original TISE  $(\hamiltonianglobal-\energyglobal)\ketstateglobal=0$ by inserting the explicit expression for $\ketsys{\statesystem(\prm)}=\braketclockglobal{\stateclock(\prm)}{\stateglobal}$ into \eqref{eq:tdse_with_effective_potential}, performing the differentiation w.r.t.~$\prm$ followed by a functional derivative $\delta/(\delta\!\bra{\stateclock})$  with respect to the  state of the environment.

\begin{figure}
    \begin{center}
    \includegraphics[width=\columnwidth]{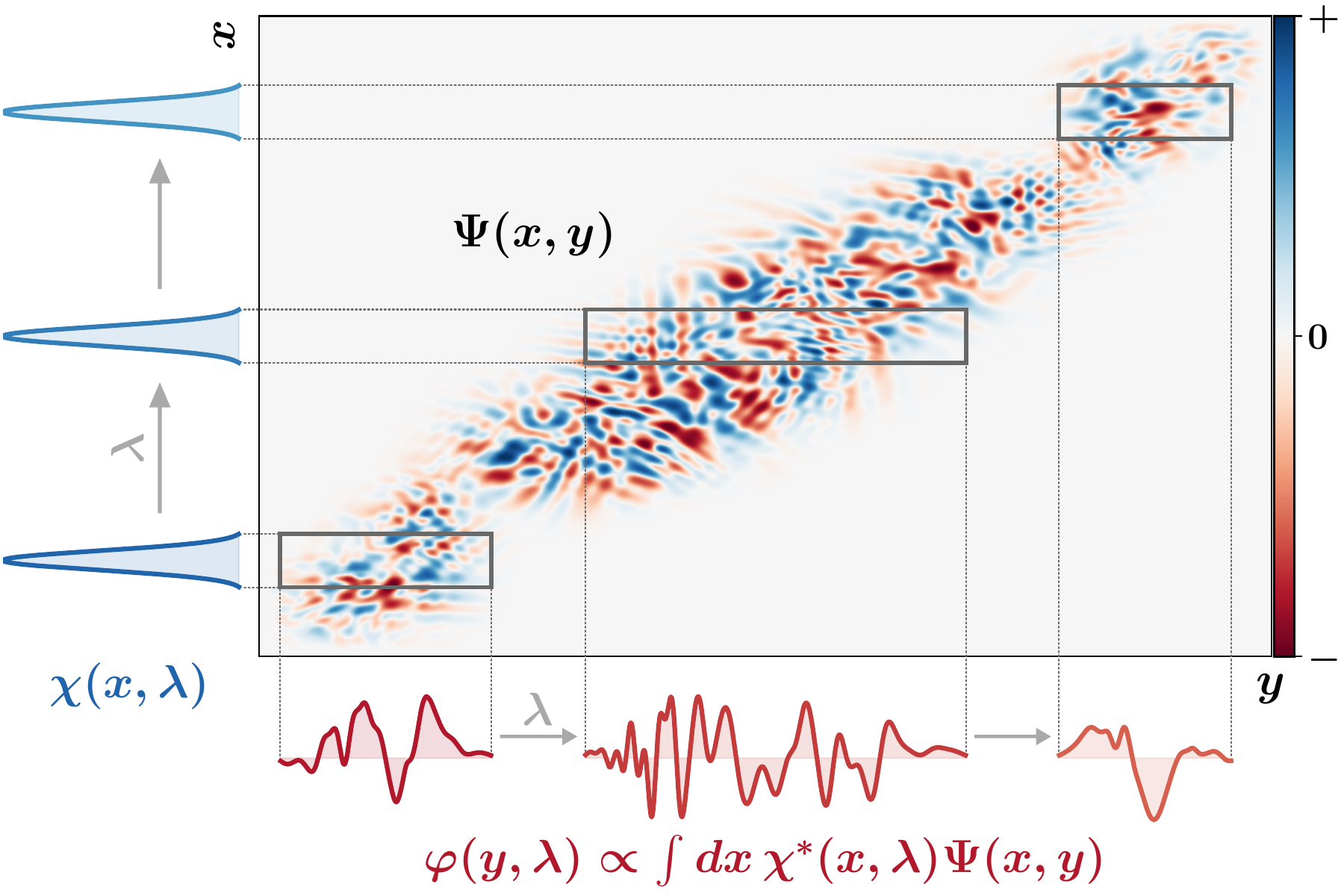}
    \caption{Emergence of system dynamics by means of the relational formalism. 
    Unitary changes in the clock state induce the system evolution through the correlations contained in the global state. 
    The invariance~\eqref{eq:global_invariance} of $\stateglobal$ ensures the concurrent system motion, which is governed by an effective clock-dependent system Hamiltonian. 
    Moreover, the entanglement in the global state admits intricate system evolutions even for relatively simple wavefunctions of the environment.}
    \label{fig:time_emergence}
    \end{center}
\end{figure}

Equation \eqref{eq:tdse_with_effective_potential} is the main result of this work and represents, to the best of our knowledge,  the first derivation of the time-dependent Schr\"odinger equation with a {\it fully general}, Hermitian time-dependent potential $\potentialsystem$ from  a static global state.
A pictorial representation of our formalism is shown in Fig.~\ref{fig:time_emergence}. 

To stay as general as possible, we have made no further assumptions regarding the interaction potential $\interaction$.
Of course, it is reasonable (although we have seen not necessary!) to assume that the interaction potential has negligible influence on the  state  $\ketclock\stateclock$ of the  environment.
Formally, this can be expressed by $[\interaction,\projectorclock]\approx 0$. Thereby, $\ketclock\stateclock$ becomes approximately an eigenstate of the interaction $\interaction$, turning $\ketclock\stateclock$  essentially into what has been described as a ``pointer state" by Zurek \cite{Zurek81}. 
Then we can write
\begin{align}
    \melclockglobal{\stateclock}{\interaction}{\stateglobal} 
    &= {\melclockglobal{\stateclock}{\projectorclock\interaction}{\stateglobal}}/{\braketclock{\stateclock}{\stateclock}}
= {\melclockglobal{\stateclock}{\interaction\projectorclock}{\stateglobal}}/{\braketclock{\stateclock}{\stateclock}}\nonumber\\
&= \frac{\melclock{\stateclock}{\interaction}{\stateclock}}{\braketclock{\stateclock}{\stateclock}}\braketclockglobal{\stateclock}{\stateglobal}=\frac{\melclock{\stateclock}{\interaction}{\stateclock}}{\braketclock{\stateclock}{\stateclock}}\ketsys{\statesystem}\,.
 \label{eq:approximation2_effective_potential}
\end{align}
The global state $\ketstateglobal$ no longer appears and renders the calculation of $\potentialsystem$ less involved.  Moreover, $\Im(\scalarpotential(\prm))= \melglobal{\stateglobal}{[\interaction, \projectorclock]}{\stateglobal}/(2iN_\prm) = 0$, which reflects the negligible influence of the interaction on the environment state.

We close with the promised concept for analytical solutions of TDSEs involving complicated, time-dependent potentials. The following, very simple example of coupled two-level systems gives a flavor for the general strategy. We consider a global Hamiltonian \eqref{eq:hamiltonian_sum} with $\hamiltoniansystem = 0$, $\hamiltonianclock = \energyclock \sigmazclock$ and the interaction $\interaction = V_0 \left( \sigmaxsystem + \sigmazsystem \right) \otimes \sigmaxclock $, where   $\sigmax,\sigmay,\sigmaz$ are the three Pauli matrices, with the additional label for system or environment.  Setting for simplicity $\energyclock = V_0\equiv 1$, we explicitly get 
\begin{align}
    \hamiltonianglobal 
    &= \begin{pmatrix*}[r]
    1 &  1 &  0 &  1 \\
    1 & -1 &  1 &  0 \\
    0 &  1 &  1 & -1 \\
    1 &  0 & -1 & -1
    \end{pmatrix*}
\end{align}
\changes{with eigenvalues $E_\pm = \pm \sqrt{3}$, where both of them are doubly degenerate. 
One eigenvector of $E_-$ in the basis $\{ \ketglobal{\upsystem \upclock} , \ketglobal{\upsystem \downclock}, \ketglobal{\downsystem \upclock}, \ketglobal{\downsystem \downclock} \}$, we take for the global state, $\stateglobal=( 1, 0, -1, -a)^{\text{T}}$, where $a=1+\sqrt{3}$.
}
Here, we use $S(\prm)=\int^\prm d\prm' \Im \scalarpotential(\prm')$ without loss of generality to simplify expressions.  With 
\begin{equation}
    \ketclock{\stateclock(\prm)} 
    = \frac{ e^{i\energyglobal_-\prm}}{2\sqrt{1+a\cos^2(\prm)}} \left[e^{-i\prm} \ketclock{\upclock} + e^{i\prm} \ketclock{\downclock} \right]
\end{equation}
we obtain from \eqref{eq:effective_potential} the effective potential
\begin{subequations}
\begin{equation}
    \potentialsystem = \vec{\interactionfunction}_S(\prm) \cdot \sigmavecsystem
\end{equation}
which enters the Schr\"odinger equation \eqref{eq:tdse_with_effective_potential}, where
\begin{align}
    \interactionfunction_{S,x} =  \interactionfunction_{S,z} &\equiv \frac{  \cos(2\prm) + a \cos^2(\prm) }{ 1 + a \cos^2(\prm) }\\
    \interactionfunction_{S,y} & \equiv  -\frac{ (a/2) \sin(2\prm) }{ 1 + a \cos^2(\prm) }\,,
\end{align}
\end{subequations}
and $\sigmavecsystem \equiv (\sigmaxsystem, \sigmaysystem, \sigmazsystem)^T$.
A physical realization would be the interaction of an electronic spin-system and a magnetic field, $\potentialsystem = - \vec{B}(\prm) \cdot \hat{\bm{\mu}}$, with  magnetic moment $\hat{\bm{\mu}} = (-e\hbar/2m_e) \sigmavecsystem$ or simply $\hat{\bm{\mu}} = -\sigmavecsystem / 2$ in atomic units.
The magnetic field has different time-dependent behavior along different directions, $\vec{B}_0 = 2[\cos(2\prm) + a \cos^2(\prm)] (\vec{e}_x + \vec{e}_z) / [1 + a \cos^2(\prm)] $ and $\vec{B}_1 = -a \sin(2\prm) \vec{e}_y /[1 + a \cos^2(\prm)] $. 

By construction, we know that the solution of the TDSE with the potential $\potentialsystem(\prm)$ is 
\begin{align}
&\ketsys{\statesystem(\prm)}\equiv  \braketclockglobal{\stateclock(\prm)}{\stateglobal}\nonumber\\
    &= \frac{ e^{i a\prm} }{ 2\sqrt{1+a\cos^2(\prm)} }  \Bigl[ \ketsys{\upsystem} - \Bigl( a\, e^{-2i\prm} + 1 \Bigr) \ketsys{\downsystem} \Bigr] \,.
\end{align}
Although the system for which we have constructed the time-dependent potential and the analytical solution of the ensuing TDSE is very simple, it admits, nevertheless, an entire class of time-dependent potentials and corresponding solutions by changing the  state $\ketclock{\stateclock(\prm)}$ of the environment.
 
Replacing the environment with a multi-level system is a straightforward extension with a semiclassical limit if the density of states of the environment in the energy interval defined by the two levels of the system becomes large. This renders the environment ``large" as compared to the system and provides a direct link between the two research strands for the emergence of time as discussed in the introduction. One can also construct a more general semiclassical limit without reference to a specific (multi-level) system with a semiclassical state $\ketclock{\stateclock(\prm)}$ from the environment and subsequent application of the stationary phase approximation, breaking implicitly the symmetry of environment and system~\cite{Gemsheim2023a}. 

While these semiclassical limits are consistent with the corresponding strand for the emergence of time, the semiclassical approach cannot uncover quantum roots of time, as we have worked them out here in form of two conditions: (i) a global state exists which respects the invariance principle \eqref{eq:global_invariance} with the global Hamiltonian and (ii) the global Hamiltonian can be decomposed  into a Hamiltonian $\hamiltoniansystem$ for the system, its   environment $\hamiltonianclock$, and their interaction $\interaction$.
With projecting the invariance principle onto an arbitrary state of the environment and all its $\prm-$dependent variants generated by ``rotating" the state with $\hamiltonianclock$, these two conditions suffice to formulate a time-dependent Schr\"odinger equation for the system with a time-dependent potential.  Thereby, we  advance the  relational approach to time by the crucial inclusion of interaction of system and environment, which so far has been possible only under very special circumstances \cite{Smith2019}. 

Since projection and separation of system and environment as well as entanglement and interaction are also  major elements of decoherence, it it is not surprising that our theory has  points of contact with Zurek's decoherence theory~\cite{Zurek03} as we have mentioned before.  However, decoherence requires time as a  prerequisite: the literal meaning of decoherence reveals it as a process {\it in} time.  The successful inclusion of interaction into the emergence of time as lined out here renders our framework suitable to ask if decoherence can be established along with emergent time in the interaction of system and environment, a question we will pursue in future work.


%


\begin{thebibliography}{0}%
\makeatletter
\providecommand \@ifxundefined [1]{%
 \@ifx{#1\undefined}
}%
\providecommand \@ifnum [1]{%
 \ifnum #1\expandafter \@firstoftwo
 \else \expandafter \@secondoftwo
 \fi
}%
\providecommand \@ifx [1]{%
 \ifx #1\expandafter \@firstoftwo
 \else \expandafter \@secondoftwo
 \fi
}%
\providecommand \natexlab [1]{#1}%
\providecommand \enquote  [1]{``#1''}%
\providecommand \bibnamefont  [1]{#1}%
\providecommand \bibfnamefont [1]{#1}%
\providecommand \citenamefont [1]{#1}%
\providecommand \href@noop [0]{\@secondoftwo}%
\providecommand \href [0]{\begingroup \@sanitize@url \@href}%
\providecommand \@href[1]{\@@startlink{#1}\@@href}%
\providecommand \@@href[1]{\endgroup#1\@@endlink}%
\providecommand \@sanitize@url [0]{\catcode `\\12\catcode `\$12\catcode
  `\&12\catcode `\#12\catcode `\^12\catcode `\_12\catcode `\%12\relax}%
\providecommand \@@startlink[1]{}%
\providecommand \@@endlink[0]{}%
\providecommand \url  [0]{\begingroup\@sanitize@url \@url }%
\providecommand \@url [1]{\endgroup\@href {#1}{\urlprefix }}%
\providecommand \urlprefix  [0]{URL }%
\providecommand \Eprint [0]{\href }%
\providecommand \doibase [0]{https://doi.org/}%
\providecommand \selectlanguage [0]{\@gobble}%
\providecommand \bibinfo  [0]{\@secondoftwo}%
\providecommand \bibfield  [0]{\@secondoftwo}%
\providecommand \translation [1]{[#1]}%
\providecommand \BibitemOpen [0]{}%
\providecommand \bibitemStop [0]{}%
\providecommand \bibitemNoStop [0]{.\EOS\space}%
\providecommand \EOS [0]{\spacefactor3000\relax}%
\providecommand \BibitemShut  [1]{\csname bibitem#1\endcsname}%
\let\auto@bib@innerbib\@empty
\end{thebibliography}%


\begin{thebibliography}{22}%
\makeatletter
\providecommand \@ifxundefined [1]{%
 \@ifx{#1\undefined}
}%
\providecommand \@ifnum [1]{%
 \ifnum #1\expandafter \@firstoftwo
 \else \expandafter \@secondoftwo
 \fi
}%
\providecommand \@ifx [1]{%
 \ifx #1\expandafter \@firstoftwo
 \else \expandafter \@secondoftwo
 \fi
}%
\providecommand \natexlab [1]{#1}%
\providecommand \enquote  [1]{``#1''}%
\providecommand \bibnamefont  [1]{#1}%
\providecommand \bibfnamefont [1]{#1}%
\providecommand \citenamefont [1]{#1}%
\providecommand \href@noop [0]{\@secondoftwo}%
\providecommand \href [0]{\begingroup \@sanitize@url \@href}%
\providecommand \@href[1]{\@@startlink{#1}\@@href}%
\providecommand \@@href[1]{\endgroup#1\@@endlink}%
\providecommand \@sanitize@url [0]{\catcode `\\12\catcode `\$12\catcode
  `\&12\catcode `\#12\catcode `\^12\catcode `\_12\catcode `\%12\relax}%
\providecommand \@@startlink[1]{}%
\providecommand \@@endlink[0]{}%
\providecommand \url  [0]{\begingroup\@sanitize@url \@url }%
\providecommand \@url [1]{\endgroup\@href {#1}{\urlprefix }}%
\providecommand \urlprefix  [0]{URL }%
\providecommand \Eprint [0]{\href }%
\providecommand \doibase [0]{https://doi.org/}%
\providecommand \selectlanguage [0]{\@gobble}%
\providecommand \bibinfo  [0]{\@secondoftwo}%
\providecommand \bibfield  [0]{\@secondoftwo}%
\providecommand \translation [1]{[#1]}%
\providecommand \BibitemOpen [0]{}%
\providecommand \bibitemStop [0]{}%
\providecommand \bibitemNoStop [0]{.\EOS\space}%
\providecommand \EOS [0]{\spacefactor3000\relax}%
\providecommand \BibitemShut  [1]{\csname bibitem#1\endcsname}%
\let\auto@bib@innerbib\@empty
\bibitem [{\citenamefont {Page}\ and\ \citenamefont
  {Wootters}(1983)}]{Page1983}%
  \BibitemOpen
  \bibfield  {author} {\bibinfo {author} {\bibfnamefont {D.~N.}\ \bibnamefont
  {Page}}\ and\ \bibinfo {author} {\bibfnamefont {W.~K.}\ \bibnamefont
  {Wootters}},\ }\bibfield  {title} {\bibinfo {title} {{E}volution without
  evolution: {D}ynamics described by stationary observables},\ }\href
  {https://doi.org/10.1103/physrevd.27.2885} {\bibfield  {journal} {\bibinfo
  {journal} {Phys. Rev. D}\ }\textbf {\bibinfo {volume} {27}},\ \bibinfo
  {pages} {2885} (\bibinfo {year} {1983})}\BibitemShut {NoStop}%
\bibitem [{\citenamefont {Wootters}(1984)}]{Wootters1984}%
  \BibitemOpen
  \bibfield  {author} {\bibinfo {author} {\bibfnamefont {W.~K.}\ \bibnamefont
  {Wootters}},\ }\bibfield  {title} {\bibinfo {title} {{"Time" replaced by
  quantum correlations}},\ }\href {https://doi.org/10.1007/bf02214098}
  {\bibfield  {journal} {\bibinfo  {journal} {International Journal of
  Theoretical Physics}\ }\textbf {\bibinfo {volume} {23}},\ \bibinfo {pages}
  {701} (\bibinfo {year} {1984})}\BibitemShut {NoStop}%
\bibitem [{\citenamefont {Giovannetti}\ \emph {et~al.}(2015)\citenamefont
  {Giovannetti}, \citenamefont {Lloyd},\ and\ \citenamefont
  {Maccone}}]{Giovannetti2015}%
  \BibitemOpen
  \bibfield  {author} {\bibinfo {author} {\bibfnamefont {V.}~\bibnamefont
  {Giovannetti}}, \bibinfo {author} {\bibfnamefont {S.}~\bibnamefont {Lloyd}},\
  and\ \bibinfo {author} {\bibfnamefont {L.}~\bibnamefont {Maccone}},\
  }\bibfield  {title} {\bibinfo {title} {Quantum time},\ }\bibfield  {journal}
  {\bibinfo  {journal} {Phys. Rev. D}\ }\textbf {\bibinfo {volume} {92}},\
  \href {https://doi.org/10.1103/physrevd.92.045033}
  {10.1103/physrevd.92.045033} (\bibinfo {year} {2015})\BibitemShut {NoStop}%
\bibitem [{\citenamefont {Marletto}\ and\ \citenamefont
  {Vedral}(2017)}]{Marletto2017}%
  \BibitemOpen
  \bibfield  {author} {\bibinfo {author} {\bibfnamefont {C.}~\bibnamefont
  {Marletto}}\ and\ \bibinfo {author} {\bibfnamefont {V.}~\bibnamefont
  {Vedral}},\ }\bibfield  {title} {\bibinfo {title} {Evolution without
  evolution and without ambiguities},\ }\bibfield  {journal} {\bibinfo
  {journal} {Phys. Rev. D}\ }\textbf {\bibinfo {volume} {95}},\ \href
  {https://doi.org/10.1103/physrevd.95.043510} {10.1103/physrevd.95.043510}
  (\bibinfo {year} {2017})\BibitemShut {NoStop}%
\bibitem [{\citenamefont {H\"{o}hn}\ \emph {et~al.}(2021)\citenamefont
  {H\"{o}hn}, \citenamefont {Smith},\ and\ \citenamefont {Lock}}]{Hoehn2021}%
  \BibitemOpen
  \bibfield  {author} {\bibinfo {author} {\bibfnamefont {P.~A.}\ \bibnamefont
  {H\"{o}hn}}, \bibinfo {author} {\bibfnamefont {A.~R.}\ \bibnamefont
  {Smith}},\ and\ \bibinfo {author} {\bibfnamefont {M.~P.}\ \bibnamefont
  {Lock}},\ }\bibfield  {title} {\bibinfo {title} {Trinity of relational
  quantum dynamics},\ }\bibfield  {journal} {\bibinfo  {journal} {Physical
  Review D}\ }\textbf {\bibinfo {volume} {104}},\ \href
  {https://doi.org/10.1103/physrevd.104.066001} {10.1103/physrevd.104.066001}
  (\bibinfo {year} {2021})\BibitemShut {NoStop}%
\bibitem [{\citenamefont {Halliwell}\ and\ \citenamefont
  {Hawking}(1985)}]{Halliwell1985}%
  \BibitemOpen
  \bibfield  {author} {\bibinfo {author} {\bibfnamefont {J.~J.}\ \bibnamefont
  {Halliwell}}\ and\ \bibinfo {author} {\bibfnamefont {S.~W.}\ \bibnamefont
  {Hawking}},\ }\bibfield  {title} {\bibinfo {title} {{Origin of structure in
  the Universe}},\ }\href {https://doi.org/10.1103/PhysRevD.31.1777} {\bibfield
   {journal} {\bibinfo  {journal} {Phys. Rev. D}\ }\textbf {\bibinfo {volume}
  {31}},\ \bibinfo {pages} {1777} (\bibinfo {year} {1985})}\BibitemShut
  {NoStop}%
\bibitem [{\citenamefont {Kucha{\v{r}}}(1992)}]{Kuchar1992}%
  \BibitemOpen
  \bibfield  {author} {\bibinfo {author} {\bibfnamefont {K.~V.}\ \bibnamefont
  {Kucha{\v{r}}}},\ }\bibfield  {title} {\bibinfo {title} {{Time and
  Interpretations of Quantum Gravity}},\ }\href
  {https://cir.nii.ac.jp/crid/1573105974273246976} {\bibfield  {journal}
  {\bibinfo  {journal} {Proc. 4th Canadian Conference on General Relativity and
  Relati vistic Astrophysics}\ } (\bibinfo {year} {1992})}\BibitemShut
  {NoStop}%
\bibitem [{\citenamefont {Kiefer}(1994)}]{Kiefer1994}%
  \BibitemOpen
  \bibfield  {author} {\bibinfo {author} {\bibfnamefont {C.}~\bibnamefont
  {Kiefer}},\ }\bibfield  {title} {\bibinfo {title} {The semiclassical
  approximation to quantum gravity},\ }in\ \href@noop {} {\emph {\bibinfo
  {booktitle} {Canonical Gravity: From Classical to Quantum}}},\ \bibinfo
  {editor} {edited by\ \bibinfo {editor} {\bibfnamefont {J.}~\bibnamefont
  {Ehlers}}\ and\ \bibinfo {editor} {\bibfnamefont {H.}~\bibnamefont
  {Friedrich}}}\ (\bibinfo  {publisher} {Springer Berlin Heidelberg},\ \bibinfo
  {year} {1994})\ pp.\ \bibinfo {pages} {170--212}\BibitemShut {NoStop}%
\bibitem [{\citenamefont {Anderson}(2012)}]{Anderson2012}%
  \BibitemOpen
  \bibfield  {author} {\bibinfo {author} {\bibfnamefont {E.}~\bibnamefont
  {Anderson}},\ }\bibfield  {title} {\bibinfo {title} {Problem of time in
  quantum gravity},\ }\href
  {https://doi.org/https://doi.org/10.1002/andp.201200147} {\bibfield
  {journal} {\bibinfo  {journal} {Annalen der Physik}\ }\textbf {\bibinfo
  {volume} {524}},\ \bibinfo {pages} {757} (\bibinfo {year}
  {2012})}\BibitemShut {NoStop}%
\bibitem [{\citenamefont {Kiefer}(2012)}]{Kiefer2012}%
  \BibitemOpen
  \bibfield  {author} {\bibinfo {author} {\bibfnamefont {C.}~\bibnamefont
  {Kiefer}},\ }\href
  {https://doi.org/10.1093/acprof:oso/9780199585205.001.0001} {\emph {\bibinfo
  {title} {{Q}uantum gravity}}},\ \bibinfo {edition} {3rd}\ ed.\ (\bibinfo
  {publisher} {Oxford University Press},\ \bibinfo {year} {2012})\BibitemShut
  {NoStop}%
\bibitem [{\citenamefont {Briggs}\ and\ \citenamefont
  {Rost}(2000)}]{Briggs2000}%
  \BibitemOpen
  \bibfield  {author} {\bibinfo {author} {\bibfnamefont {J.~S.}\ \bibnamefont
  {Briggs}}\ and\ \bibinfo {author} {\bibfnamefont {J.~M.}\ \bibnamefont
  {Rost}},\ }\bibfield  {title} {\bibinfo {title} {Time dependence in quantum
  mechanics},\ }\href {https://doi.org/10.1007/s100530050554} {\bibfield
  {journal} {\bibinfo  {journal} {The European Physical Journal D}\ }\textbf
  {\bibinfo {volume} {10}},\ \bibinfo {pages} {311} (\bibinfo {year}
  {2000})}\BibitemShut {NoStop}%
\bibitem [{\citenamefont {Briggs}\ and\ \citenamefont
  {Rost}(2001)}]{Briggs2001}%
  \BibitemOpen
  \bibfield  {author} {\bibinfo {author} {\bibfnamefont {J.~S.}\ \bibnamefont
  {Briggs}}\ and\ \bibinfo {author} {\bibfnamefont {J.~M.}\ \bibnamefont
  {Rost}},\ }\bibfield  {title} {\bibinfo {title} {{On the Derivation of the
  Time-Dependent Equation of {S}chr{\"o}dinger}},\ }\href
  {https://doi.org/10.1023/a:1017525227832} {\bibfield  {journal} {\bibinfo
  {journal} {Foundations of Physics}\ }\textbf {\bibinfo {volume} {31}},\
  \bibinfo {pages} {693} (\bibinfo {year} {2001})}\BibitemShut {NoStop}%
\bibitem [{\citenamefont {Schild}(2018)}]{Schild2018}%
  \BibitemOpen
  \bibfield  {author} {\bibinfo {author} {\bibfnamefont {A.}~\bibnamefont
  {Schild}},\ }\bibfield  {title} {\bibinfo {title} {Time in quantum mechanics:
  {A} fresh look at the continuity equation},\ }\href
  {https://link.aps.org/doi/10.1103/PhysRevA.98.052113} {\bibfield  {journal}
  {\bibinfo  {journal} {Phys. Rev. A}\ }\textbf {\bibinfo {volume} {98}},\
  \bibinfo {pages} {052113} (\bibinfo {year} {2018})}\BibitemShut {NoStop}%
\bibitem [{\citenamefont {Chataignier}(2022)}]{Chataignier2022}%
  \BibitemOpen
  \bibfield  {author} {\bibinfo {author} {\bibfnamefont {L.}~\bibnamefont
  {Chataignier}},\ }\bibfield  {title} {\bibinfo {title} {Beyond semiclassical
  time},\ }\href {https://doi.org/10.1515/zna-2022-0106} {\bibfield  {journal}
  {\bibinfo  {journal} {Zeitschrift f\"{u}r Naturforschung A}\ }\textbf
  {\bibinfo {volume} {77}},\ \bibinfo {pages} {805} (\bibinfo {year}
  {2022})}\BibitemShut {NoStop}%
\bibitem [{\citenamefont {Gemsheim}(2023)}]{Gemsheim2023}%
  \BibitemOpen
  \bibfield  {author} {\bibinfo {author} {\bibfnamefont {S.}~\bibnamefont
  {Gemsheim}},\ }\emph {\bibinfo {title} {The emergence of time with
  interactions in quantum and classical mechanics}},\ \href@noop {} {Ph.D.
  thesis},\ \bibinfo  {school} {Technische Universit{\"a}t Dresden} (\bibinfo
  {year} {2023})\BibitemShut {NoStop}%
\bibitem [{\citenamefont {Gemsheim}\ and\ \citenamefont
  {Rost}(2023{\natexlab{a}})}]{Gemsheim2023b}%
  \BibitemOpen
  \bibfield  {author} {\bibinfo {author} {\bibfnamefont {S.}~\bibnamefont
  {Gemsheim}}\ and\ \bibinfo {author} {\bibfnamefont {J.~M.}\ \bibnamefont
  {Rost}},\ }\bibfield  {journal} {\bibinfo  {journal} {in preparation}\
  }\href@noop {} {} (\bibinfo {year} {2023}{\natexlab{a}})\BibitemShut
  {NoStop}%
\bibitem [{Note1()}]{Note1}%
  \BibitemOpen
  \bibinfo {note} {\protect \leavevmode {\protect Since the clock
  projection includes an identity operation on the system part, it is fully
  described by $(\protect \hat {\protect \mathds {1}}_{\protect \text
  {S}}\otimes \bra {\chi }) \ket *{\Psi }\protect \tmspace -\thinmuskip
  {.1667em}\rangle = \DOTSB \sum@ \slimits@ _{kl} c_{kl} (\protect \hat
  {\protect \mathds {1}}_{\protect \text {S}}\ket {\varphi _k}_{\protect \text
  {S}} ) \otimes \braket {\chi }{\chi _l} = \DOTSB \sum@ \slimits@ _k \ket
  {\varphi _k}_{\protect \text {S}} (\DOTSB \sum@ \slimits@ _l c_{kl} \braket
  {\chi }{\chi _l}) \in \protect \mathcal {H}_{\protect \text {S}}$ for a
  representation of $\ket *{\Psi }\protect \tmspace -\thinmuskip
  {.1667em}\rangle $ in the arbitrary global basis $\{ \ket *{\varphi _k
  \otimes \chi _l}\protect \tmspace -\thinmuskip {.1667em}\rangle \}_{kl}$. We
  will omit those identity operators where obvious in favor of a better
  readability}.}\BibitemShut {Stop}%
\bibitem [{\citenamefont {Zurek}(2003)}]{Zurek03}%
  \BibitemOpen
  \bibfield  {author} {\bibinfo {author} {\bibfnamefont {W.~H.}\ \bibnamefont
  {Zurek}},\ }\bibfield  {title} {\bibinfo {title} {Decoherence, einselection,
  and the quantum origins of the classical},\ }\href
  {https://doi.org/10.1103/RevModPhys.75.715} {\bibfield  {journal} {\bibinfo
  {journal} {Rev. Mod. Phys.}\ }\textbf {\bibinfo {volume} {75}},\ \bibinfo
  {pages} {715} (\bibinfo {year} {2003})}\BibitemShut {NoStop}%
\bibitem [{\citenamefont {Callender}(2017)}]{Callender2017}%
  \BibitemOpen
  \bibfield  {author} {\bibinfo {author} {\bibfnamefont {C.}~\bibnamefont
  {Callender}},\ }\href {https://doi.org/10.1093/oso/9780198797302.001.0001}
  {\emph {\bibinfo {title} {{What Makes Time Special?}}}}\ (\bibinfo
  {publisher} {Oxford University Press},\ \bibinfo {year} {2017})\BibitemShut
  {NoStop}%
\bibitem [{\citenamefont {Smith}\ and\ \citenamefont
  {Ahmadi}(2019)}]{Smith2019}%
  \BibitemOpen
  \bibfield  {author} {\bibinfo {author} {\bibfnamefont {A.~R.~H.}\
  \bibnamefont {Smith}}\ and\ \bibinfo {author} {\bibfnamefont
  {M.}~\bibnamefont {Ahmadi}},\ }\bibfield  {title} {\bibinfo {title}
  {Quantizing time: Interacting clocks and systems},\ }\href
  {https://doi.org/10.22331/q-2019-07-08-160} {\bibfield  {journal} {\bibinfo
  {journal} {Quantum}\ }\textbf {\bibinfo {volume} {3}},\ \bibinfo {pages}
  {160} (\bibinfo {year} {2019})}\BibitemShut {NoStop}%
\bibitem [{\citenamefont {Zurek}(1981)}]{Zurek81}%
  \BibitemOpen
  \bibfield  {author} {\bibinfo {author} {\bibfnamefont {W.~H.}\ \bibnamefont
  {Zurek}},\ }\bibfield  {title} {\bibinfo {title} {Pointer basis of quantum
  apparatus: Into what mixture does the wave packet collapse?},\ }\href
  {https://doi.org/10.1103/PhysRevD.24.1516} {\bibfield  {journal} {\bibinfo
  {journal} {Phys. Rev. D}\ }\textbf {\bibinfo {volume} {24}},\ \bibinfo
  {pages} {1516} (\bibinfo {year} {1981})}\BibitemShut {NoStop}%
\bibitem [{\citenamefont {Gemsheim}\ and\ \citenamefont
  {Rost}(2023{\natexlab{b}})}]{Gemsheim2023a}%
  \BibitemOpen
  \bibfield  {author} {\bibinfo {author} {\bibfnamefont {S.}~\bibnamefont
  {Gemsheim}}\ and\ \bibinfo {author} {\bibfnamefont {J.~M.}\ \bibnamefont
  {Rost}},\ }\bibfield  {title} {\bibinfo {title} {From quantum to
  semiclassical emergence of time},\ }\bibfield  {journal} {\bibinfo  {journal}
  {in preparation}\ }\href@noop {} {} (\bibinfo {year}
  {2023}{\natexlab{b}})\BibitemShut {NoStop}%
\end{thebibliography}
\end{document}